\documentclass[conference,comsoc,10pt]{IEEEtran}

\usepackage[utf8]{inputenc}
\usepackage{graphicx}
\graphicspath{ {./images/} }
\usepackage{amsmath}
\usepackage{mathtools}
\usepackage{graphicx} 
\usepackage{comment}
\usepackage{caption,subcaption}
\usepackage{float}
\usepackage{xcolor} 
\usepackage{soul} 
\usepackage{comment}
\usepackage{amssymb}
\usepackage{graphicx, float}
\usepackage{caption,lipsum}
\usepackage{dblfloatfix}
\usepackage{blindtext}
\usepackage{algorithm}
\usepackage{algorithmic}
\usepackage{amsmath}
\usepackage{booktabs}
\usepackage{tabularx}
\usepackage{makecell}
\usepackage{verbatim}
\usepackage{comment}
\usepackage{amsmath}
\usepackage{colortbl}
\usepackage{bm}
\usepackage{enumitem}
\DeclareMathOperator*{\argmax}{arg\,max}

\ifCLASSOPTIONcompsoc
  \usepackage[nocompress]{cite}
\else
  \usepackage{cite}
\fi
\ifCLASSINFOpdf
\else
\fi
\begin{document}
\title{Multi-User Scheduling in Hybrid Millimeter Wave Massive MIMO Systems }

\author{Seyedeh~Maryam~Hosseini,~Shahram~Shahsavari,
        and~Catherine~Rosenberg\\
        Department of Electrical and Computer Engineering, University of Waterloo, Waterloo, ON, Canada.\\
Emails: \{seyedeh.maryam.hosseini, shahram.shahsavari, cath\}@uwaterloo.ca}

\maketitle

\begin{abstract}
While mmWave bands provide a large bandwidth for mobile broadband services, they suffer from severe path loss and shadowing. Multiple-antenna techniques such as beamforming (BF) can be applied to compensate the signal attenuation. We consider a special case of hybrid BF called \textit{per-stream hybrid BF} (PSHBF) which is easier to implement than the general hybrid BF because it circumvents the need for joint analog-digital beamformer optimization. Employing BF at the base station enables the transmission of multiple data streams to several users in the same resource block. In this paper, we provide an offline study of proportional fair multi-user scheduling in a mmWave system with PSHBF to understand the impact of various system parameters on the performance. We formulate multi-user scheduling as an optimization problem. To tackle the non-convexity, we provide a feasible solution and show through numerical examples that the performance of the provided solution is very close to an upper-bound. Using this framework, we provide extensive numerical investigations revealing several engineering insights.
\end{abstract}

\IEEEpeerreviewmaketitle

\section{Introduction}
Massive multi-input multi-output (MIMO) technology has shown a great potential to meet the ever growing demand for high spectral and energy efficiencies in cellular networks. Millimeter wave (mmWave) bands (between 30 GHz and 300 GHz) have also shown a great potential to offer high data rates. However, the characteristics of mmWave bands are challenging, in particular, the path loss and shadowing is more intense compared to microwave bands.

Beamforming (BF) is a key enabler of wireless communications in mmWave systems. While fully digital BF (DBF) has been widely recommended for microwave massive MIMO systems \cite{marzetta2016fundamentals}, its implementation is costly and highly power consuming in mmWave bands \cite{hybrid-survey} due to having a separate RF chain per antenna.  \cite{khalili-adc,shokri-2015,khalili-adc2}. To reduce deployment cost and power consumption of DBF, several hybrid BF (HBF) architectures have been proposed where the BF processing is divided between based-band digital and RF analog domains \cite{hybrid-survey}. These architectures use a large number of antennas with much fewer RF chains at the BS and at the UEs. Thus, HBF is more cost and power efficient than DBF in mmWave bands and is also capable of supporting multiple concurrent data streams to serve multiple UEs in the same physical resource block (PRB)~\cite{shokri-2015} using  spatial division multiple access (SDMA).

In this paper,  we consider a special case of hybrid BF called \textit{per-stream hybrid BF} (PSHBF) where each data stream is connected to a separate baseband processing.  Each of the $L$ data streams is processed separately in the digital baseband domain first and then is transferred to the analog domain using $K'$ RF chains where $K' = K/L$. The analog processing in PSHBF  is implemented by a network of phase shifters connected to the antenna array with $N$ elements. The purpose of digital processing in PSHBF is to enable the modification of the amplitude of the BF coefficients as it is impractical to perform it in the analog RF domain due to the implementation cost. The downside of separating digital processing for different streams is the lack of control on the inter-stream interference in the baseband. 
However, this interference can also be controlled by the user scheduler by selecting UEs, that will be activated in a given PRB, appropriately. One of the major practical advantages of PSHBF is that full channel state matrix is not required to compute the digital and analog beamformers. Rather, the BF is efficiently implemented by adopting
feedback-assisted BF codebooks as suggested by \cite{love-2015}. 


Proper radio resource management (RRM) is essential to enable high throughput in SDMA-based mmWave systems. In this paper, we focus on multi-user scheduling given PSHBF architecture. We note that several prior works (e.g., \cite{hegde2018joint,he2017joint,kwon2016joint,jiang2016joint,bogale2015user}) have investigated multi-user scheduling problem in MIMO systems with general HBF architecture but from the best of our knowledge, no such study exists for PSHBF. Additionally, the existing studies considering general HBF have several common limitations. For example, while discrete modulation and coding schemes (MCSs) are used in practice in digital communication systems, it has not been considered in most of the prior multi-user scheduling studies. Rather, Shannon capacity formula, i.e., $\log_2(1 + SINR)$, is commonly adopted. However, this formula only provides an upper-bound for the achievable data rate and cannot capture all properties of the discrete MCSs. Consequently, using Shannon formula may cause performance degradation (for instance by selecting the wrong subset of UEs) or misleading insights especially when it is used for user scheduling. For instance, there is a minimum SINR threshold below which the data rate is zero when using discrete MCSs. However, a UE always has a positive achievable data rate when using Shannon formula, which may lead to a sub-optimal user scheduling. Furthermore, there is a maximum data rate when using discrete MCSs but  any SINR increase translates into a data rate improvement when adopting Shannon formula. Additionally, most of the state of the art studies considered one PRB or time slot of the system. As a result, such studies are typically focused on short-term performance metrics such as UE SINRs and data rates and do not consider longer term performance metrics such as UE throughput and fairness. Another limitation in several prior works is assuming single antenna UEs which restricts the performance in mmWave scenarios. We consider multiple antenna UEs in this paper to combat with path loss.

In this paper, we consider the downlink of a multi-user massive MIMO small cell with PSHBF at the BS and the UEs. We provide an offline study to investigate multi-user scheduling, i.e., the selection of a subset of UEs for each PRB, the distribution of power to the selected UEs and the selection of an MCS per selected UE. Furthermore, unlike several existing studies in the literature where heuristic approaches are proposed, in this offline study, we intend to provide theoretical bounds to investigate the impact of several system parameters on the performance. We first formulate the proportional fair (PF) multi-user scheduling problem in a OFDM system incorporating PSHBF, different system parameters, and discrete MCSs. This problem is non-linear and non-convex. We present a convex relaxed version of the problem solving which provides a performance upper-bound for the original problem. Additionally, we derive a feasible solution for the original problem by processing the optimal solution to the relaxed problem. We show through numerical evaluations that the performance of the feasible solution is close to the upper-bound obtained from the relaxed problem, implying that we have solved the original problem to quasi optimality. Subsequently, we use the developed multi-user scheduling optimization framework  to investigate the impact of various system parameters on the performance, leading to several engineering insights.

\section{System Model} \label{sec:sysmodel}
We consider the downlink  of a single small cell multi-user massive MIMO system operating in a mmWave band. We assume that the BS is equipped with $N_{b}$ antennas and $K_{b}$ RF chains, and can transmit a maximum of $L$ data streams simultaneously. Additionally, we let $\mathcal{U}$ and $U$ denote the set and the number of UEs, respectively. We also assume that UE $u$ has $N_{u}$ antennas and $K_{u}$ RF chains. 

\subsection{Per-stream Hybrid MIMO Architecture}

\begin{figure}[H]
     \centering
     \begin{subfigure}[b]{0.4\textwidth}
         \centering
         \includegraphics[width=\textwidth]{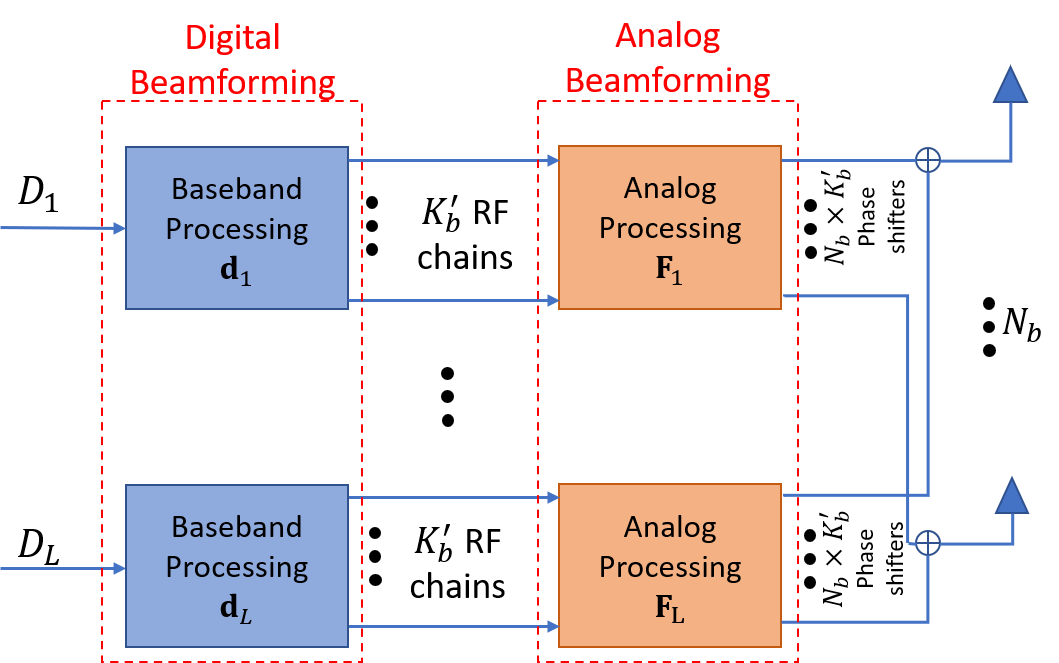}
         \caption{BS}
         \label{fig:beamform-bs}
     \end{subfigure}
     \hfill
     \begin{subfigure}[b]{0.4\textwidth} \centering
         \includegraphics[width=\textwidth]{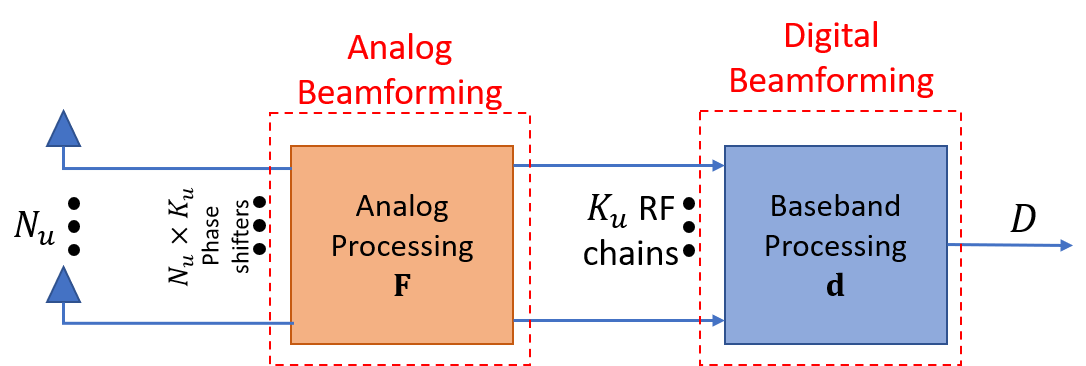}
         \caption{UE}
         \label{fig:beamform-ue}
     \end{subfigure}
     \hfill
      \caption{The considered per-stream hybrid BF architecture at (a) BS and  (b) UE.}
        \label{fig:beamform2}
\end{figure}

We consider the per-stream hybrid MIMO architecture for BS and each UE depicted in Fig.~\ref{fig:beamform2}, where there is a digital baseband processing followed by an analog RF processing for each data stream. While we explain the architecture using the BS parameters, we consider the same architecture for the UEs but with different parameters. Baseband processor $i$ multiplies data stream $D_i$ by a complex vector $\mathbf{d}_i \in \mathbb{C}^{K'_b}$. The output of the baseband processor is then fed into an analog processor using $K'_b$ RF chains. We consider a fully connected PSHBF architecture where the analog processor is composed of $K'_b\times N_b$ phase shifters, i.e., there is a phase shifter between each RF chain and each antenna in the antenna array. 
The functionality of the analog processing unit corresponding to data stream $D_i$ is represented by matrix $\mathbf{F}_i = [\mathbf{f}_{i,1},...,\mathbf{f}_{i,K_b^{'}}]\in \mathbb{C}^{N_b \times K_b'}$, where $\mathbf{f}_{i,n}$ is an analog BF vector in the form of
$\mathbf{f}_{i,n} =  \frac{1}{N_{b}} [e^{j\phi_1^{i,n}},\ldots, e^{j\phi_{N_{b}}^{i,n}}]^T$, where $\phi_j^{i,n}$ is the amount of phase shift applied to the path connecting the $n$th output of the digital processor of $D_i$ to the $j$th antenna. 
As a result, the overall BF vector applied to data stream $D_i$ is $\mathbf{w}_i = \mathbf{F}_i \mathbf{d}_i$. Without loss of generality we assume that BF vectors are normalized, i.e., $\left\|\mathbf{F}_i \mathbf{d}_i \right\| ^2 =1$.

In this structure, the baseband and analog BF units are combined to enable directional transmissions. The architecture is characterized by three parameters, the total number of RF chains $K_b$, the maximum number of streams $L$ and the number of RF chains per stream $K'_b$,  such that $K_b=L\times K'_b$. 
We note that the separation of baseband and analog allows for modifying both amplitude and phase of the input signals in a practical fashion as it is not easy and cost effective to tune the amplitudes in the RF analog domain. We assume that each UE is only allowed to receive one stream  i.e. $K_u=K'_u$.


\subsection{Beam Alignment}
We assume that BS and UE $u$ have predefined BF codebooks $\mathcal{C}_b$ and $\mathcal{C}_u$, respectively. We define:
\begin{align*}
    &\mathcal{C}_b=\left\{\mathbf{w}_j \in \mathbb{C}^{N_b} :  \left\| \mathbf{w}_j \right\|^2=1,   j=1,...,2^{B_b}\right\},\\
    &\mathcal{C}_u=\left\{\mathbf{v}_j \in \mathbb{C}^{N_u} :  \left\| \mathbf{v}_j \right\|^2=1,   j=1,...,2^{B_u}\right\}, \forall u\in \mathcal{U},
\end{align*}
where $2^{B_b}$ and $2^{B_u}$ are the number of BF codewords in $\mathcal{C}_b$ and $\mathcal{C}_u$, respectively. Prior to the data transmission, we assume that a round of beam alignment (BA) occurs to enable the use of directional beams with high BF gains during data transmission. The BA is a procedure to determine the best BF codeword pair to connect the BS to each UE. Note that since we assume each UE can receive at most one data stream, it is sufficient to choose only one pair of BF codewords per-UE; one to be applied at that UE and one to be applied at the BS when transmitting to that UE. While there are several approaches for BA, we adopt a well-know approach called \textit{exhaustive search} and  find the  beam pair $(\mathbf{w}_u^*,\mathbf{v}_u^*)$ leading to the highest SNR for UE $u$ \cite{exhaustive}. 

\subsection{Frame Structure and Operational Processes}

\begin{figure}[t]
    \centering
    \includegraphics[scale=0.5]{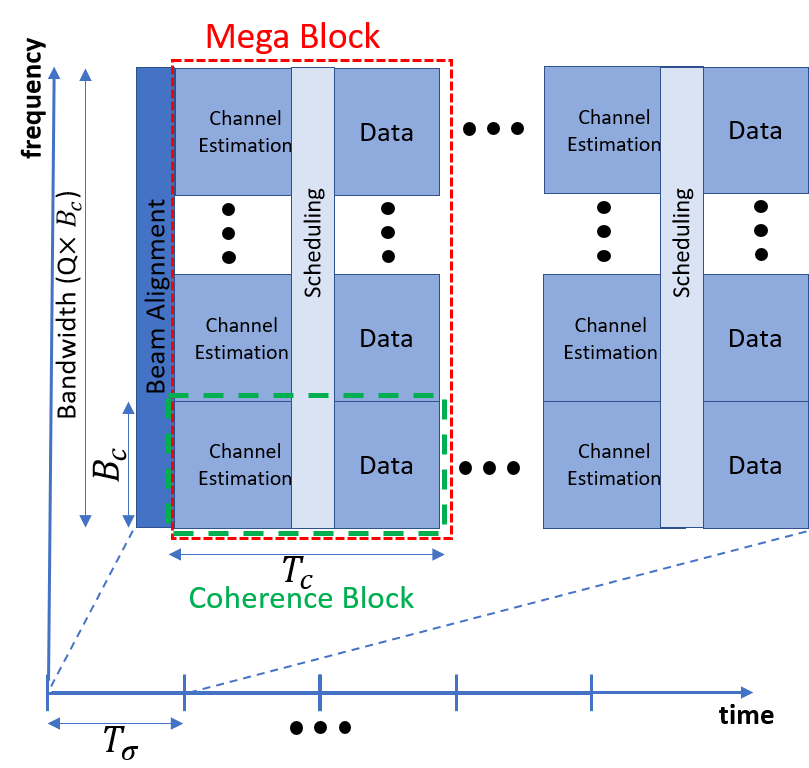}
    \caption{Overview of the processes in the time-frequency domain including beam alignment, channel estimation, and data transmission.}
    \label{block}
\end{figure}
In this section, we elaborate on the structure of the time-frequency frame of interest and explain the  processes related to RRM, occurring during the system operation.

We represent the MIMO channel between the BS and each UE by a complex $N_u \times N_b$ matrix which will be defined later. This matrix incorporates small-scale fading coefficients, angles of arrival (AoAs) of the channel spatial clusters at UE, angles of departure (AoDs) of the spatial channel clusters at the BS, as well as large-scale fading coefficients (modeling path loss and shadowing). The small-scale fading part is assumed to be constant over a coherence time interval of length $T_c$ as well as over a coherence bandwidth of length $B_c$ in frequency. Furthermore, we assume that large-scale fading coefficients, AoAs, AoDs are constant over time intervals of length $T_{\sigma}$ and are independent of the frequency within the system bandwidth. Typically, $T_{\sigma}>>T_c$.
We assume that BA is performed at the beginning of every $T_\sigma$ interval. The BA information is used to reduce the overhead of channel estimation (CE) as the channel coefficients are only estimated in the directions obtained in BA as will be explained later.

The channel matrix is constant over a $T_c$ [sec]$\times B_c$ [Hz] time-frequency block, called \textit{coherence block} (CB). We assume that the system bandwidth, denoted by $BW$, can be divided into $Q$ coherence bandwidth intervals, i.e., $BW=QB_c$. We define a mega block (MB) as the group of $Q$ coherence blocks in the frequency domain as depicted in Fig.~\ref{block}. Therefore, the time and frequency spans of each MB are $T_c$ and $BW$, respectively. In each MB, the downlink operation is performed in three phases (note that BA is done less frequently as discussed above). The first phase is CE which is performed at the beginning of each CB. In the second phase, the estimated CSI is used to perform  multi-user scheduling for all the remaining PRBs in all $Q$ CBs. In the third phase, transmissions to the selected UEs are performed based on the computed scheduling. We note that the reason behind defining a MB is that the scheduling is performed jointly for all CBs `vertically' contained in the MB to improve diversity and fairness. 
We let $T_{\sigma}=N_o T_c$, where $N_o$ is the number of MBs within each $T_{\sigma}$ interval. Let $\mathbf{H}_{q,u,o}\in \mathbb{C}^{N_u \times N_b}$ denote the channel matrix between the BS and UE $u$ in CB $q$ within MB $o$. For the sake of brevity, hereafter we drop the MB index. 

After BA, the BS knows which BF vector to use for transmission to UE $u$ and UE $u$ knows which BF vector to use for reception if scheduled. Note that the beams to use for a given UE are the same for all CBs. Let $\mathbf{w}_u^*$ and $\mathbf{v}_u^*$ denote the corresponding BF vectors to be applied at the BS and UE $u$ if it is scheduled. At the beginning of each CB, the effective signal and interference channels are estimated for each UE. Let $h_{q,u,u}^\text{eff} \in \mathbb{C}$ denote the effective signal  channel between UE $u$ and the BS in any PRB of CB $q$. We have $h_{q,u,u}^\text{eff}=\left({\mathbf{v}^*_u}\right)^H \mathbf{H}_{q,u}\left(\mathbf{w}_u^*\right)$. Furthermore, let $h_{q,u,n}^\text{eff} \in \mathbb{C}$ denote the effective interference channel between the BS and UE $u$ due to the BS transmission to UE $n\not=u$ in any PRB of CB $q$. We define $h_{q,u,n}^\text{eff}=\left({\mathbf{v}^*_u}\right)^H \mathbf{H}_{q,u}\left(\mathbf{w}_n^*\right)$. We assume that the effective signal channel and interference channels are estimated for each UE prior to the scheduling (see Fig.~\ref{block}). The channel estimation can be done without a large overhead. Let $\mathcal{C}_b^\star$ be the set of indices of all distinct BS beams in the set $\{\mathbf{w}_u^*\}_{u\in\mathcal{U}}$ which is the output of BA. Each member of $\mathcal{C}_b^\star$ is the index of a BF codeword at the BS which has been selected at least for one UE during BA. One efficient approach for estimation of effective channels is for each UE to set its reception BF codeword to the one selected during the BA (i.e., $\mathbf{v}^*_u$ for UE $u$) and then the BS transmits a pilot signal with each BF codeword whose index belongs to $\mathcal{C}_b^\star$. As a result, UE $u$ can estimate the required effective signal and interference channels by processing the received pilot signals. While the size of the BS codebook can be large, we note that the size of $\mathcal{C}_b^\star$ does not exceed the number of UEs within the cell which is typically not large in small mmWave cells. This would limit the overhead of the effective channel estimation process. Furthermore, since the BS is capable of sending multiple streams, it can use different frequency subchannels to send multiple pilots at the same time. This way the UEs can estimate multiple effective channels at the same time reducing the estimation overhead. In this paper, we ignore the estimation error and assume that the effective channels can be estimated perfectly. After effective channel estimation, the scheduler receives the set of estimated effective channel coefficients and selects a subset of UEs for each PRB in the remaining part of each CB.
Next, we elaborate on the scheduling problem.

\section{Multi-user Scheduling} \label{sec:scheduling}
\begin{figure}[t]
    \centering
    \includegraphics[width=0.4\textwidth]{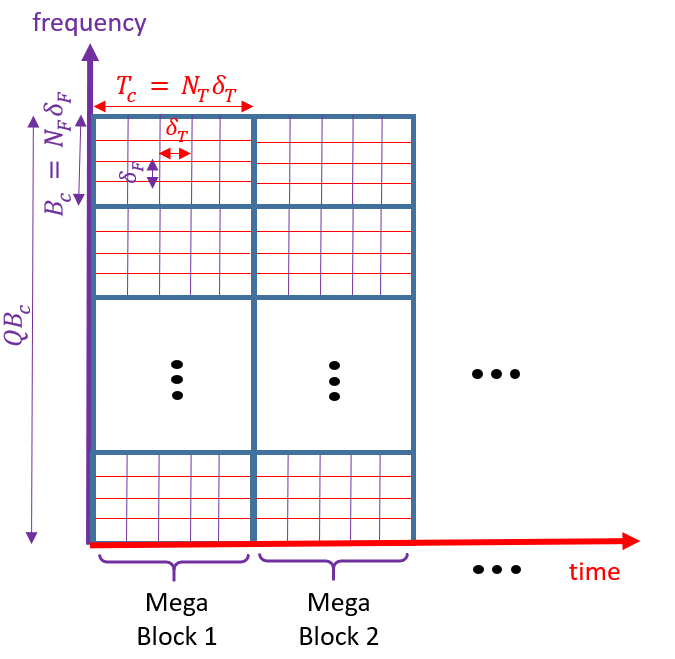}
    \caption{Structure of data transmission blocks  consisting of multiple PRBs.}
    \label{fig:SchedulingExample}
\end{figure}

We consider the OFDM structure depicted in Fig.~\ref{fig:SchedulingExample}. Note that the figure does not show the overhead of BA and CE and the processing time required for scheduling. This figure only depicts the data blocks shown in the Fig.~\ref{block}. The time span of a PRB is denoted by $\delta_T$ and its frequency span is $\delta_F$. Let $T_c = N_T\delta_T$, where $N_T$ is the number of time slots within each CB, be the time span of a MB. Additionally, let $B_c = N_F \delta_F$, where $N_F$ is the number of frequency subchannels within each CB, be the frequency span of a MB. Therefore, each time slot includes $N_F$ vertical PRBs for DL data per CB and each CB has $N_T N_F$ PRBs for DL data in total. Having the results of BA and perfect channel estimation, we want to design a scheduling algorithm in a given MB that selects  a subset of UEs for each PRB.

\subsection{Problem Formulation}
Recall that $\mathcal{C}_b^\star$ is the set of indices of all distinct beams in the set $\{\mathbf{w}_u^*\}_{u\in\mathcal{U}}$ which is the output of BA. We can think of the scheduling process as follows: for each time slot, we select a subset of $L' \leq L$ BS beams that can be activated in the corresponding PRBs of that time slot and then select a subset of UEs for each PRB to be served by the selected beams. It is not feasible to change the subset of active beams over frequency subchannels at the same time slot. Note that the PSHBF architecture employed at the BS can create at most $L$ beams at a time. Let $\mathcal{L}$ be the set of all possible distinct BS beam subsets of size $L'\leq L$, i.e., 
\begin{align}
    \mathcal{L}=\Big\{ \ell: &\ell = \{b^\ell_1, \ldots,b^\ell_{L'}\}, b^\ell_j \in \mathcal{C}_b^\star,1\leq j \leq L', L' \leq L \Big\}. \notag
\end{align}
The set $\mathcal{L}$ is constructed after BA  based on $\mathcal{C}_b^\star$.
Note that it is feasible and maybe better to have less than $L$ active beams in a time slot to reduce intra-cell interference or to increase the transmit power per beam.

Assuming that a beam set $\ell $ is selected for a given time slot, the next step is to select the set of UEs which are going to be served in each PRB of that time slot. Clearly, if set $\ell$ is selected for a time slot, then a UE whose preferred BS beam index (i.e., the index of the beam corresponding to the BF vector $\mathbf{w}^*_u$) is not in set $\ell$, will not be scheduled at that time slot. However, as a BS beam can be the preferred beam for multiple UEs, there may be multiple sets of UEs which could be scheduled in a time slot given beam set $\ell$ is selected. Therefore, for each beam set $\ell=\{b_1^\ell \ldots, b_{L'}^\ell \}$, we define $\mathcal{M}_\ell$ as the set of UE tuples that could be scheduled using the BS beams in $\ell$. We have 
$$\mathcal{M}_\ell=\Big\{z_{\ell}: z_{\ell}= (d_1, \ldots,d_{L'}), d_j \in \mathcal{U},   b(d_j)=b_j^\ell,j \leq L' \Big\}$$ where $b(d_j)$ is the index of the preferred BS beam for UE $d_j$, i.e., the index of the BS beam corresponding to the BF vector $\mathbf{w}^*_{d_j}$.  $\mathcal{M}_\ell$ is constructed for every beam set $\ell \in \mathcal{L}$ after BA. Note that $\mathcal{M}_\ell$ is not computed frequently since BA is not performed  often.

Our scheduling optimization problem  will determine $\alpha_\ell$ which is the fraction of time slots in which a beam set $\ell$ is selected in a given MB. It is important to note that as the channel coefficients in each CB within a MB do not change, it does not make any difference where in time within a MB a beam set $\ell$ is selected as long as the fraction $\alpha_\ell$ is preserved. 
Clearly we have 
\begin{align}
&\sum_{\ell \in \mathcal{L}} \alpha_\ell =1, \label{eq:alphax1}\\
&0 \leq \alpha_\ell \leq 1, & \forall \ell \in \mathcal{L}, \label{eq:alphax2}\\
&N_T\alpha_\ell \in \mathbb{Z},  & \forall \ell \in\mathcal{L}. \label{eq:alphax3}
\end{align}
The last constraint is because we have $N_T$ time slots within each MB and the number of time slots assigned to a beam set should be an integer.

Given the BS selected beam set  $\ell$, various UE tuples can be scheduled in each of the $QN_F$ vertical PRBs of each time slot in a MB. Let 
 $\beta^\ell_{q}(z_\ell)$ be the fraction of PRBs within CB $q$ that  UE tuple $z_{\ell}$ is scheduled given that $\ell$ is active. Note that as the channel is constant in each CB, it does not make any difference which PRBs are allocated to which UE tuple as long as the fractions are preserved. 
 We have 
\begin{align}
&\sum_{z_{\ell} \in \mathcal{M}_\ell} \beta^\ell_{q}(z_\ell) = 1, & \forall \ell \in \mathcal{L}, \forall q\in \mathcal{Q}, \label{eq:betax1}\\
&0 \leq \beta^\ell_{q}(z_\ell) \leq 1  & \forall \ell \in \mathcal{L}, \forall q\in \mathcal{Q}, \forall  z_{\ell} \in \mathcal{M}_\ell, \label{eq:betax2}\\ 
&N_F\beta^\ell_{q}(z_\ell) \in \mathbb{Z} & \forall \ell \in \mathcal{L}, \forall q\in \mathcal{Q}, \forall z_{\ell} \in \mathcal{M}_\ell, \label{eq:betax3}
\end{align} 
where $\mathcal{Q} = \{1,2,,\ldots,Q\}$ is the set of vertical CB indices within a MB. (\ref{eq:betax3}) is due to the fact that we have $N_F$ PRBs at each time slot within each CB and the number of PRBs assigned to a UE group should be integer.

Given that beam set $\ell$ is active in a time slot, we assume that the BS divides its transmit power, denoted by $P_{BS}$, equally among the beams in this set. Furthermore, we assume that the power assigned to each beam is uniformly distributed over the $QN_F$ vertical PRBs.  Let $\gamma^{q,u}_{\ell}(z_\ell)$ denote the SINR of UE $u$, if beam set $\ell$ and UE tuple $z_{\ell}$ are active at any PRB of CB $q$. We have $\forall \ell \in \mathcal{L},\forall q \in \mathcal{Q}, \forall z_\ell \in \mathcal{M}_\ell, \forall u \in \mathcal{U},$
\begin{align}
&\gamma^{q,u}_{\ell}(z_\ell) =
\begin{cases}
     &\frac{ \left| h_{q,u,u}^{\text{eff}} \right|^2 P_{\text{BS}}/(|\ell|QN_F)}{\sum\limits_{\substack{n\in z_{\ell}\\n\neq u}} \left| h_{q,u,n}^{\text{eff}} \right|^2P_{\text{BS}}/(|\ell|QN_F) + \sigma^2_{\text{PRB}}}, \;\; \text{if} \;\; u\in z_{\ell}, \\
    & 0, \;\; \text{otherwise},
\end{cases}\notag
\end{align}
where $\sigma^2_{\text{PRB}}$ is the noise power for one PRB and  $\sigma^2_{\text{PRB}}=\delta_FN_0$, where $N_0$ is the noise power spectral density.

Let $r_\ell^{q,u}(z_\ell)= f(\gamma^{q,u}_{\ell}(z_\ell))$ be the spectral efficiency that UE $u$ would experience in a PRB of CB $q$ if beam set $\ell$ and UE tuple $z_\ell$ are selected, where, $f(\cdot)$ is the step-wise non-decreasing function, translating SINR to spectral efficiency according to the set of available discrete MCSs. As the transmit powers are determined in advance, given the effective channel estimates, one can compute the SINRs and the corresponding spectral efficiencies prior to the scheduling. This reduces the complexity of the scheduling problem.

We can define the throughput of a UE (in bits per second) in a MB as 
\begin{align}
    \lambda_u = B_c\sum_{\ell \in \mathcal{L}} \alpha_\ell \sum_{\substack{z_{\ell} \in \mathcal{M}_\ell\\ 
        }} \sum_{q\in\mathcal{Q}} \beta^\ell_{q}(z_\ell) r_\ell^{q,u}(z_\ell), \quad \forall u\in \mathcal{U} \label{eq:lamda}.
\end{align}
Let $R_u$ be the average throughput that UE $u$ has received in the past $W$ MBs up to the start of the current MB. We want to be proportionally fair over $W+1$ MBs. Hence, we consider $\Delta(u)= WR_u+\lambda_u$ as
the utility function for UE $u$ at the beginning of the current MB.
Given $\mathcal{U}$, $\mathcal{L}$, $L$, $\mathcal{M}_\ell$, $R_u$, $W$, $N_T$, $N_F$, and \{$r^{q,u}_{\ell}(z_\ell)\}$, we formulate the proportional fair multi-user scheduling optimization problem $\boldsymbol{\Pi}$ at a MB as:
\begin{align}
    & \boldsymbol{\Pi}:~ \max_{\beta^\ell_{q}(z_\ell), \alpha_\ell} \sum_{u \in \mathcal{U}} \log(WR_u+\lambda_u) \notag\\
    & \text{s.t.} \;\; (\ref{eq:alphax1}), (\ref{eq:alphax2}), (\ref{eq:alphax3}), (\ref{eq:betax1}), (\ref{eq:betax2}), (\ref{eq:betax3}), \;  \text{and} \; (\ref{eq:lamda}) \nonumber
\end{align}

Problem $\boldsymbol{\Pi}$ is  non-convex, non-linear, and includes integer constraints. To tackle this problem, we first relax the problem by removing the integer constraints. Additionally, we introduce the following change of variables where $\xi^\ell_{q}(z_\ell)\triangleq\alpha_\ell\beta^\ell_{q}(z_\ell)$ to get rid of the product of the optimization variables in \eqref{eq:lamda}. Let $\boldsymbol{\Pi}_r$ denote the relaxed transformed problem:
\begin{align}
    & \boldsymbol{\Pi}_r:~ \max_{\xi^\ell_{q}(z_\ell), \alpha_\ell} \sum_{u \in \mathcal{U}} \log(WR_u+\lambda_u) \notag\\
    & \text{s.t.} \;\; \sum_{\ell \in \mathcal{L}} \alpha_\ell =1, 
~0 \leq \alpha_\ell \leq 1,\forall \ell\in\mathcal{L},\notag \\
    & \lambda_u = B_c\sum_{\ell \in \mathcal{L}}  \sum_{\substack{z_{\ell} \in \mathcal{M}_\ell \\ 
        }} \sum_{q=1}^{Q} \xi^\ell_{q}(z_\ell) r^{q,u}_{\ell}(z_\ell), \qquad \forall u \in \mathcal{U}, \\
    &\sum_{z_{\ell} \in \mathcal{M}_\ell} \xi^\ell_{q}(z_\ell) \leq \alpha_\ell, \; \qquad \qquad \qquad \forall \ell\in \mathcal{L} , \forall q\in \mathcal{Q},\\
    &0 \leq \xi^\ell_{q}(z_\ell) \leq 1, \; \quad \qquad \forall \ell \in \mathcal{L},\forall q \in \mathcal{Q}, \forall z_\ell \in \mathcal{M}_\ell.
\end{align}
The variables are all continuous in Problem~$\boldsymbol{\Pi}_r$. Furthermore, it can be shown that this problem is convex. Hence Problem~$\boldsymbol{\Pi}_r$ can be solved efficiently by solvers such as MINOS in AMPL \cite{fourer1987ampl}. Additionally, the optimal objective of Problem~$\boldsymbol{\Pi}_r$ is an upper-bound for that of Problem~$\boldsymbol{\Pi}$. 

Solving Problem~$\boldsymbol{\Pi}_r$ provides the optimal values $\alpha^*_\ell$ and ${\xi_q^{\ell}}^*(z_\ell)$ for this problem. We define ${\beta_q^{\ell}}^*(z_\ell)=\alpha^*_\ell/{\xi_q^{\ell}}^*(z_\ell)$. If $\alpha^*_\ell$ and ${\beta_q^{\ell}}^*(z_\ell)$ are feasible for Problem~$\boldsymbol{\Pi}$ (i.e., if they satisfy constraints \eqref{eq:alphax3} and \eqref{eq:betax3}), then they are the optimal solution of the original problem,  otherwise, we use the heuristic given in Algorithm~\ref{alg:feasible-solution} below to build a feasible solution for Problem~$\boldsymbol{\Pi}$. In this algorithm, we use simple rounding techniques to find $\alpha'_\ell$ and ${\beta'}^\ell_{q}(z_\ell)$ satisfying constraints \eqref{eq:alphax3} and \eqref{eq:betax3}, respectively.   
Also, we  use the optimal objective value of Problem~$\boldsymbol{\Pi}_r$ as a benchmark to evaluate the quality of the feasible solution. We update $R_u$ as $R_u := (1-1/W)R_u + \lambda_u/W$ at the end of each MB.

\begin{algorithm} 
\caption{Feasible Solution Construction for Problem  $\boldsymbol{\Pi}$}
\begin{algorithmic}[1]
    \STATE Solve Problem $\boldsymbol{\Pi}_r$ and find $\alpha_\ell, \forall \ell \in \mathcal{L}$ and $\xi^\ell_{q}(z_\ell),  \forall \ell \in \mathcal{L}, \forall  z_\ell \in \mathcal{M}_\ell, \forall q\in \mathcal{Q}$
    \STATE Compute $\beta^\ell_{q}(z_\ell) = \xi^\ell_{q}(z_\ell)/\alpha_\ell,  \forall \ell \in \mathcal{L}$ and $\xi^\ell_{q}(z_\ell),  \forall \ell \in \mathcal{L}, \forall  z_\ell \in \mathcal{M}_\ell, \forall q\in \mathcal{Q}$
    \FOR{$\ell=1,\ldots,\mathcal{L}$}
        \STATE $\alpha'_\ell \leftarrow \lfloor{N_T \times \alpha_\ell} \rfloor/N_T$ 
        \FOR{$q\in \mathcal{Q}$}
        \STATE $\beta'^\ell_{q}(z_\ell) \leftarrow \lfloor{N_F \times \beta^\ell_{q}(z_\ell) \rfloor}/N_F, \quad \forall z_\ell\in \mathcal{M}_\ell$
        \ENDFOR
    \ENDFOR
    \IF{$(\sum_{\ell\in \mathcal{L}}\alpha'_\ell  < 1)$} 
        \STATE  $\ell^\star \leftarrow \argmax_{l}{(\alpha_\ell)}$
        \STATE    $\alpha'_{\ell^\star} \leftarrow \alpha'_{\ell^\star} +(1 - \sum_{\ell\in \mathcal{L}}\alpha'_\ell)$
    \ENDIF
    \FOR {$q \in \mathcal{Q}$}
    \FOR {$\ell \in \mathcal{L}$}
        \IF{$(\sum_{z^\ell \in \mathcal{M}_\ell} \beta'^\ell_{q}(z_\ell)  < 1)$}
    \STATE    $z_{\ell}^\star \leftarrow \argmax_{z_\ell}{(\beta'^\ell_{q}(z_\ell))}$ \\
    \STATE $\beta'^{\ell}_{q}(z_{\ell}^\star) \leftarrow \beta'^{\ell}_{q}(z_{\ell}^\star) + (1 - \sum_{z^\ell \in \mathcal{M}_\ell}\beta'^\ell_{q}(z_\ell))$
        \ENDIF
    \ENDFOR
    \ENDFOR
    \end{algorithmic}
    \label{alg:feasible-solution}
\end{algorithm}

\begin{table*}[htbp]
\caption{Simulation Parameters}
\begin{center}
\scalebox{1}{
\begin{tabular}{llll}
\toprule
 \textbf{{Parameter}} & \textbf{{Value}} & \textbf{{Parameter}} & \textbf{{Value}} \\
\midrule
  Number of antennas at BS $N_b$ & \{128, 64\} 
 & Number of antennas at UE $N_u$ & \{16, 8\} \\
\midrule
  Number of RF chains at BS $K_b$ & \{2,4,8,16\} 
 & Number of RF chains at UE $K_u$ & \{1,4\} \\
\midrule
  Number of RF chains per beam at BS $K'_b$ & \{1,2,4\}  
 & Number of RF chains per beam at UE $K'_u$ & \{1,4\} \\
\midrule
  BF codebook size at BS $2^{B_b}$ & \{16,32\} 
 & BF codebook size at UE $2^{B_u}$ & \{4,8\} \\
\midrule
  Transmit power at BS $P_{BS}$ & 30 dBm 
 & Carrier frequency  $f_c$,  & 28 GHz\\  
 \midrule
 System bandwidth $BW$ & 200 MHz
 & Noise power spectral density $N_0$ & -174 dBm/Hz  \\
 \midrule
   Coherence time $T_c$ & 5 msec
 & Coherence bandwidth $B_c$ & 8.64 MHz  \\

\midrule
  PRB time duration $\delta _T$ &  0.125 msec 
 & PRB bandwidth $\delta _F$ & 720 KHz \\
\midrule
  Number of PRBs per coherence time $N_T$ & 40
 & Number of PRBs per coherence bandwidth $N_F$ & 12    \\
\midrule  

  Number of CBs in frequency $Q$  & 22 
 & Moving average parameter $W$ & 100 \\
\bottomrule
\end{tabular}}
\label{tab1}
\end{center}
\end{table*}

\section{Numerical Results }
\label{sec:sim}
In this section, we provide numerical results to address our research questions about the impact of various system parameters on the performance. 
We consider the DL of a small cell of radius $r=75$ m within which the UEs are distributed uniformly. The BS height is set to 10 m and the BS power budget is considered to be $P_{BS} = 30$~dBm. We assume that BS and UEs are equipped with uniform linear antenna arrays (ULAs) with half-wavelength inter-element spacing. The system operates at $28$~GHz and the bandwidth is assumed to be $200$~MHz. Furthermore, we assume $Q=22$ leading to a coherence bandwidth of $B_c = 8.64$~MHz. Additionally, we assume $T_c=5$~msec. Such values for $B_c$ and $T_c$ are typical for low-mobility scenarios in mmWave frequency bands \cite{Du2017}. We adopt the set of discrete MCSs from \cite[Table 7.2.3-1]{3gpp-rate}. The rest of the simulation parameters are listed in Table \ref{tab1}.

Next, we  provide the well-known channel model used in our numerical examples. Subsequently, we elaborate on the BF codebook design approaches used in our numerical results. We end this section with several numerical examples that illustrate the impact of various system parameters on the performance

\subsection{Channel Model}
We consider a narrow-band multi-ray geometric channel model that incorporates multiple scattering clusters, each including a number of spatial paths \cite{Rappaport-2014}. It has been adopted in several prior studies and has been experimentally validated in mmWave bands \cite{Rappaport-2014}. 
According to this model, the MIMO channel between BS and UE $u$ at CB $q$ of an MB is defined as the following $N_{u} \times N_{b}$ matrix\footnote{For the brevity of notation, we continue to focus on one MB and remove the MB index.}:
\begin{equation}
    \mathbf{H}_{q,u}=\frac{1}{\sqrt{N_\text{path}}}\sum_{d=1}^{N_{\text{cluster}}} \sum_{l=1}^{N_\text{path}} g_{d,l}^{q,u}\mathbf{a}_{u}(\phi_{d,l}^{u})\mathbf{a}_{b}^\text{H}(\phi_{d,l}^{b}), \label{eq:channel}
\end{equation}
where $N_\text{cluster}$ and $N_\text{path}$ are the number of clusters and the number of  spatial paths in each cluster, respectively. Furthermore, $g^{q,u}_{d,l}$ represents the fading coefficient corresponding to the spatial path $l$ belonging to cluster $d$. It is assumed that $g_{d,l}^{q,u} = \sigma_d^u \kappa^{q,u}_{d,l}$ where $\kappa^{q,u}_{d,l} \sim \mathcal{CN}(0,1)$ models small-scale fading and $\sigma_d^u$ models large-scale fading. We have $\sigma_d^u = \sqrt{(\nu_d^u 10^{-0.1 PL_u})}$ where $\nu_d^u$ is the fraction of the power carried by the $d$-th cluster and $PL_u$ (in dB) represents the impact of omnidirectional  path loss and shadowing for UE $u$. Moreover, $\phi_{d,l}^{b}$ and $\phi_{d,l}^{u}$  in \eqref{eq:channel} denote, respectively,  the AoD and the AoA corresponding to the $l^{th}$ path  of cluster $d$ of the channel. Note that the large-scale channel parameters such as path loss, AoAs, and AoDs do not depend on the CB index $q$. We refer the reader to \cite{Rappaport-2014} where the typical values of the channel parameters are given for 28 GHz mmWave band. Additionally, $\mathbf{a}_{b}(.)$ and $\mathbf{a}_{u}(.)$ are the \textit{array response vectors} at the BS and UE $u$, respectively. The array response vector depends on the antenna array type. For instance, for a ULA including $N$ antennas with half-wavelength inter-element distance, $\mathbf{a}(.)$ is given by $\mathbf{a}(\phi)=[1, e^{j\pi sin(\phi)}, \ldots, e^{j(N-1)\pi sin(\phi)}]^T$, where $\phi$ is the AoD if the array is at the transmitter and it is AoA if the array is at the receiver.

\subsection{Beamforming Codebook Design}
We adopt the BF codebook design method proposed in \cite{love-2015}  to generate BF codebook for PSHBF. The objective of this method is to generate a set of BF vectors corresponding to  non-overlapping beams covering the whole azimuth plane while having as flat as possible BF gains in the coverage region of each beam. The inputs of this algorithm is the number of antennas (e.g., $N_b$ at the BS), the number of BF codewords in the BF codebook (e.g., $2^{B_b}$ at the BS), the number of RF chains per beam (e.g., $K'_b$ described in Fig.~\ref{fig:beamform2}). The method proposed by \cite{love-2015} uses a smart exhaustive search algorithm to solve an optimization problem that has two parameters, named $M$ and $T$ in \cite{love-2015} which are set to $15$ and $3$, respectively, in our simulations. It is important to note that the quality of the generated beams depends on these parameters.


\subsection{Performance Evaluation}
Let $n(U)$ be a network realization obtained by generating $U$ uniformly distributed UE positions in the cell excluding a ring with radius of 6 m around the BS. It also includes the channel parameters corresponding to these positions. For a given set of system parameters and a value of $U$, we generate $Z=100$ random realizations and provide results averaged over the realizations. For each realization, we perform BA once in the beginning and run the scheduler for $100$ consecutive MBs. For each MB, we solve the upper-bound problem $\boldsymbol{\Pi}_r$ using the MINOS package in AMPL\cite{fourer1987ampl}. 
Subsequently, we compute the feasible  throughput $\lambda'_u(n)$ for UE $u$ in realization $n$ which is averaged over $100$~MBs\footnote{Note that we can obtain per-UE feasible throughput for each MB using the feasible values $\alpha'_\ell$ and ${\beta'}_q^{\ell}(z_\ell)$ (obtained from Algorithm~\ref{alg:feasible-solution}) in \eqref{eq:lamda}. The average throughput for each UE can be computed by averaging per-UE throughput over all MBs.}. 
We characterize the performance by the geometric mean (GM) of the UE feasible throughput,  a measure of both system throughput and fairness, which is computed for each realization $n$ as $GM(n)=\big[\prod_{u=1}^{U} \lambda'_u(n)\big]^{1/U}$.
Finally, we take the arithmetic average of the GMs over all realizations and compute $\overline{GM}=\sum_n GM(n)/Z$. 

\begin{figure}
    \centering
    \includegraphics[width=0.45\textwidth]{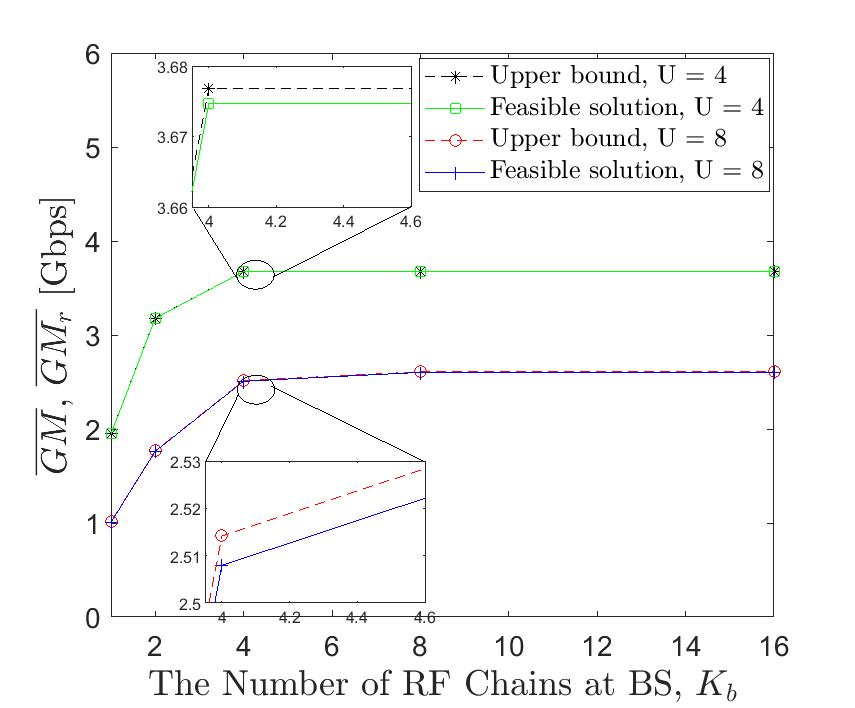}
    \caption{Comparing the upper bounds and the feasible solutions when $2^{B_u}=4, 2^{B_b}=32, N_u=8, N_b=128$, $K'_b=1$, $K_u=K'_u = 1$.}
    \label{fig:zoom}
\end{figure}
 
\subsubsection{Quality of the feasible solution}
To investigate the quality of the feasible solution, we define $GM_r(n)=\big[\prod_{u=1}^{U} \lambda''_u(n)\big]^{1/U}$, where $\lambda''_u(n)$ is obtained from the solution of relaxed Problem~$\boldsymbol{\Pi}_r$. We also define $\overline{GM}_r=\sum_n GM_r(n)/Z$. Clearly, $\overline{GM}_r$ is an upper-bound for $\overline{GM}$.
We consider a scenario where the numbers of antennas at each UE and the BS are $N_u=8$ and $N_b=128$, respectively, and the BF codebooks consist of $2^{B_u}=4$ and $2^{B_b}=32$ beams at each UE and  BS, respectively. The BF codebooks are designed assuming $K'_b=K'_u=1$. 
Fig.~\ref{fig:zoom} depicts $\overline{GM}_r$ and $\overline{GM}$ as a function of the number of  RF chains at the BS, $K_b$, for $U\in\{4,8\}$ while considering a single RF chain per UE, i.e., $K_u=1$. We observe that the upper-bounds and their corresponding feasible solutions are very close (maximum $0.3\%$ gap over different values of $K_b$ and $U$) implying that the feasible solution is quasi optimal. We have observed a similar trend for several other choices of system parameters. Consequently, we only focus on the feasible solution obtained by Algorithm~\ref{alg:feasible-solution} in the remainder of this section.

\begin{figure}
     \centering
     \begin{subfigure}[b]{0.41\textwidth}
         \centering
         \includegraphics[width=\textwidth]{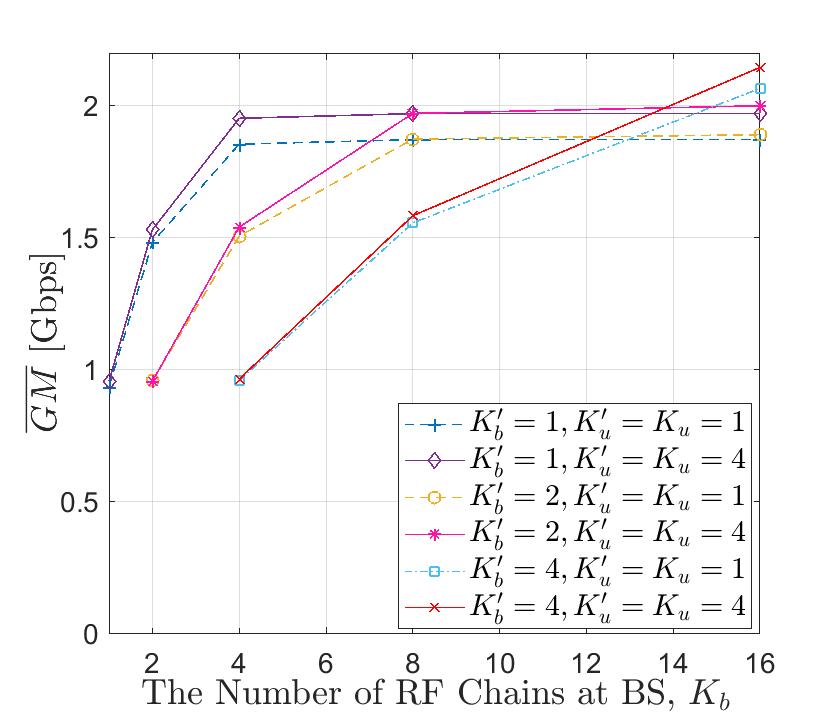}
         \caption{GM  versus the number of RF chains.}
         \label{fig:k-prime-1}
     \end{subfigure}
     \hfill
     \begin{subfigure}[b]{0.41\textwidth} \centering
         \includegraphics[width=\textwidth]{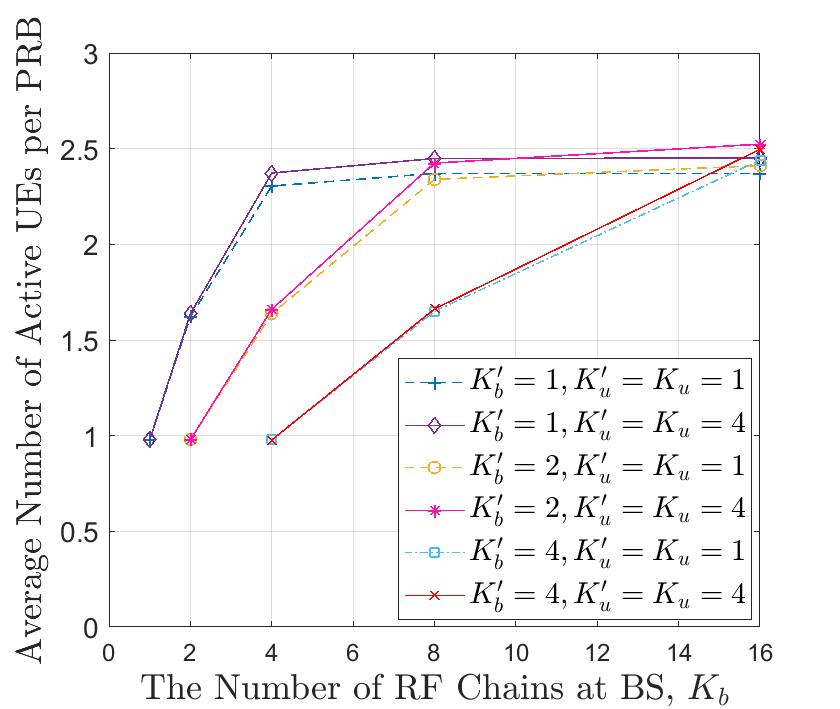}
         \caption{The number of active UEs per PRBs}
         \label{fig:k-prime-2}
     \end{subfigure}
     \hfill
        \caption{The impact of $K'_b$, $K'_u$ on the performance when $U=8$, $N_b=128$, $N_u=16$, $2^{B_b}=16$, $2^{B_u}=4$.}
        \label{fig:k-prime}
\end{figure}
\subsubsection{Impact of system parameters} 
Next, we study the impact of system parameters on the system performance. In the first example, we focus on  $K_b$, $K'_b$, and $K'_u$. To this end, we fix $U=8$, $N_b=128$, $N_u=16$, $2^{B_b}=16$, and $2^{B_u}=4$. Fig.~\ref{fig:k-prime-1} and Fig.~\ref{fig:k-prime-2} show $\overline{GM}$ and the average number of active UEs per PRB, respectively, as a function $K_b$ for various scenarios. 


While having a larger $K'_b$ leads to a higher quality in BF codebook design \cite{love-2015}, we observe from Fig.~\ref{fig:k-prime-1} that increasing $K'_b$ reduces $\overline{GM}$ substantially when $K_b$ is small. The reason is that given a fixed $K_b$, increasing $K'_b$ restricts the maximum number of data streams $L$ which reduces the average number of active UEs per PRB as shown in Fig.~\ref{fig:k-prime-2}. Therefore, unless there is a sufficiently large number of RF chains at the BS, increasing $K'_b$ deteriorates the performance significantly by limiting $L$. Furthermore, while increasing the number of streams $L$ (equivalent to increasing $K_b$ for a fixed $K'_b$) is very advantageous initially, it is not beneficial after a certain point where the performance reaches a plateau.  Consequently,  in practice, if building a system with a few RF chains, they should be used to increase the number of data streams $L$ by taking $K'_b=1$. Given a larger number of RF chains, the benefits of increasing $L$ is marginal and the extra RF chains can be used to increase the number of RF chains per data stream $K'_b$.  

Comparing the cases with $K_u=K'_u=1$ and $K_u=K'_u=4$, we can see that increasing $K'_u$ is always beneficial since the UE has only one incoming stream and increasing $K'_u$ always improves the quality of beam shapes applied at the UEs at the cost of using more RF chains. However, the improvement caused by increasing $K'_u$ is not significant and it may not be worthwhile to increase the number of RF chains at UE if there is a single incoming data stream at the UE. 

\begin{figure}[t]
     \centering
     \begin{subfigure}[b]{0.42\textwidth}
         \centering
         \includegraphics[width=\textwidth]{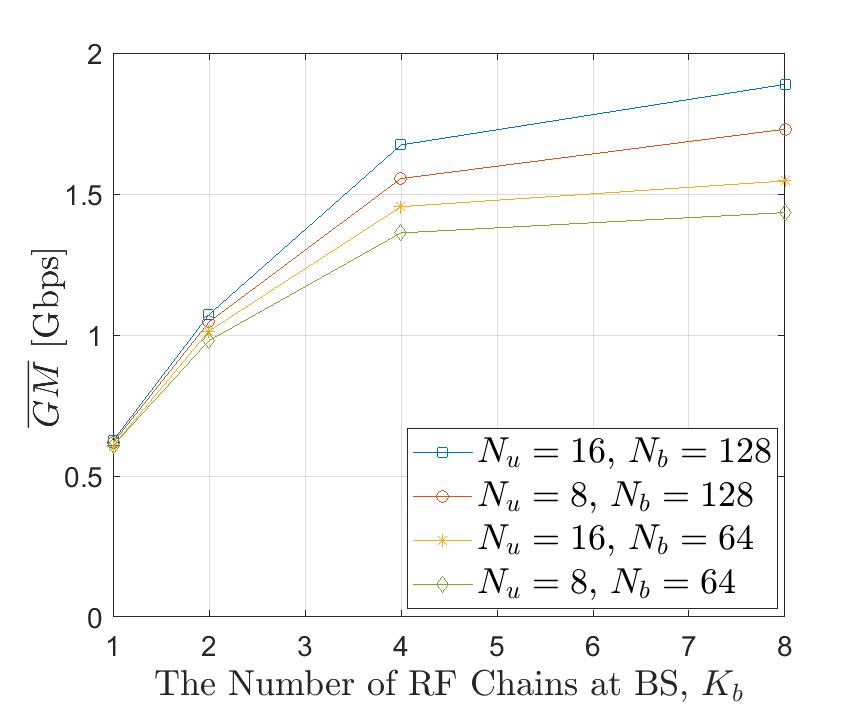} 
         \caption{$\overline{GM}$ versus the number of RF chains at BS.}
         \label{fig:impact-antenna-1}
     \end{subfigure}
     \hfill
     \begin{subfigure}[b]{0.42\textwidth} \centering
         \includegraphics[width=\textwidth]{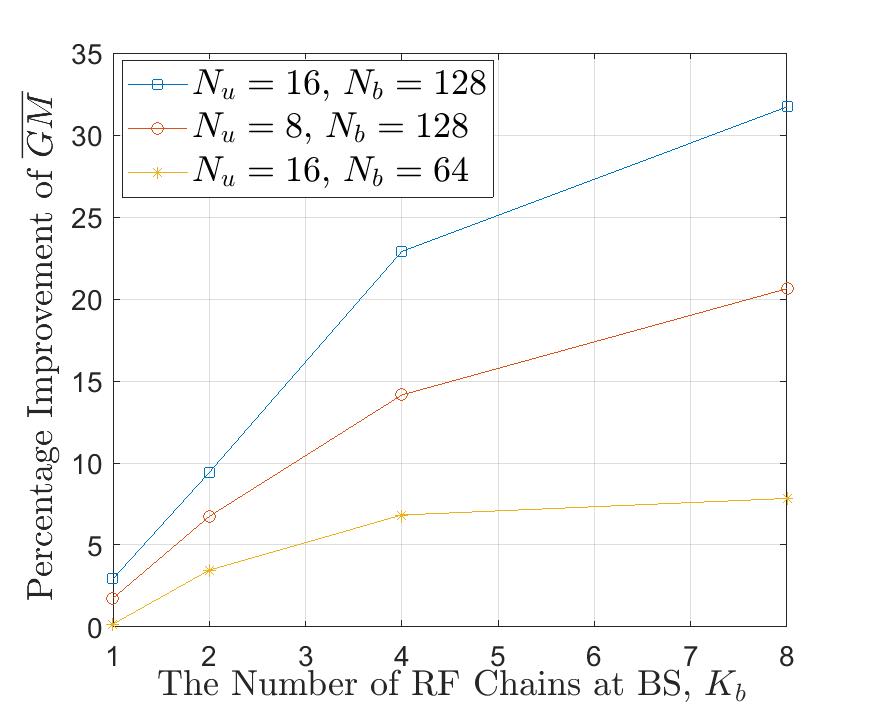}
         \caption{The $\overline{GM}$ improvement.}
         \label{fig:impact-antenna-2}
     \end{subfigure}
     \hfill
        \caption{Impact of the number of antennas on the performance when $U=12$, $2^{B_u}=4, 2^{B_b}=32, K'_u=1, K'_b=1$}
        \label{fig:impact-antenna}
\end{figure}


In the next example, we fix $U=12$, $K'_u=K'_b=1$, $2^{B_u}=4$ and $2^{B_b}=32$ and change the number of antennas at the BS and UEs. Fig.~\ref{fig:impact-antenna-1} shows $\overline{GM}$ as a function of $K_b$ for different values of the number of antennas at each UE and the BS. The first observation is that doubling the number of antennas at the BS or UE is not impacting $\overline{GM}$ significantly when the total number of RF chains $K_b$ is small at the BS. Given the number of BF codewords generated by Method~1, increasing the number of antennas improves the beam pattern quality as discussed earlier. However, a small $K_b$ restricts the number of active UEs per PRB. Hence the interference is limited and improving the beam shapes is only beneficial through improving the average BF gain due to having better beam patterns. However, the performance improvement due to the larger BF gain is marginal as the BS power is not split among a large number of UEs and is high enough to provide UEs with high spectral efficiencies. 
Another important observation from Fig.~\ref{fig:impact-antenna-1} is that increasing the total number of RF chains at the BS increases the impact of the number of antennas at the BS and UE on the performance. To elaborate, Fig.~\ref{fig:impact-antenna-2} depicts the percentage improvement in $\overline{GM}$ with respect to the baseline case with $N_u=8$ and $N_b=64$ (green curve with diamond markers in Fig.~\ref{fig:impact-antenna-1}), as a function of $K_b$. We observe that increasing the number of antennas at the BS or UE is more advantageous for larger values of $K_b$. The reason is that, given a BF codebook, increasing the number of antennas improves the beam shapes (by providing more degrees of freedom) reducing the interference among the beams which is more important when there are more RF chains at the BS since lower inter-beam interference can be exploited to activate a larger subset of UEs per PRB.

\begin{figure}[t]
     \centering
     \begin{subfigure}[b]{0.42\textwidth}
         \centering
         \includegraphics[width=\textwidth]{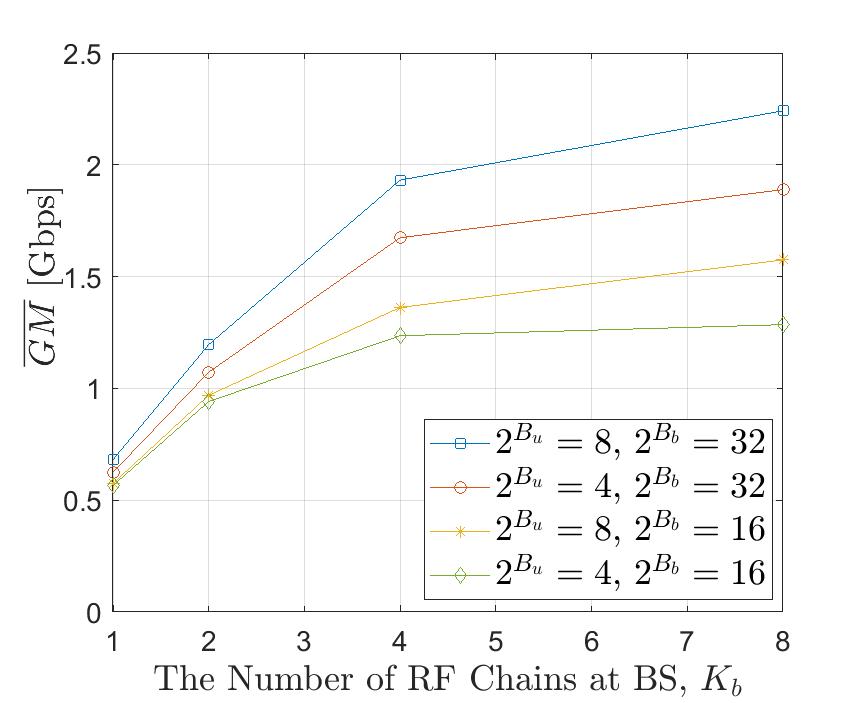}
         \caption{$\overline{GM}$ versus the number of RF chains at BS.}
         \label{fig:impact-beam-1}
     \end{subfigure}
     \hfill
     \begin{subfigure}[b]{0.42\textwidth} \centering
          \includegraphics[width=\textwidth]{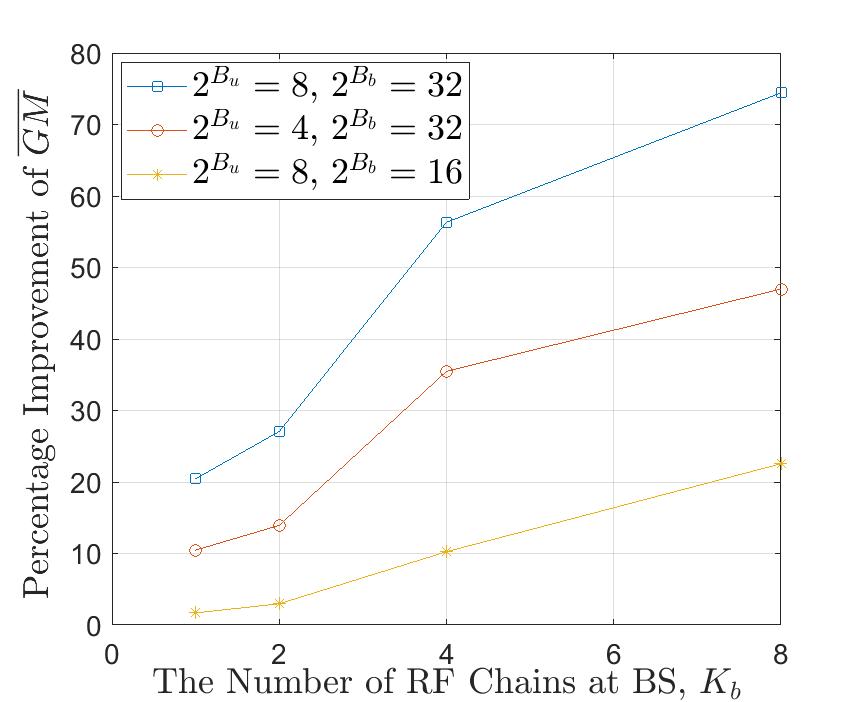}
         \caption{The $\overline{GM}$ improvement.}
         \label{fig:impact-beam-2}
     \end{subfigure}
     \hfill
        \caption{Impact of the BF codebook size on the performance when $U=12$, $N_b=128, N_u=16, K'_u=1, K'_b=1$.}
        \label{fig:impact-beam}
\end{figure}

Next, we investigate the impact of the number of beams belonging to the UE and BS BF codebooks on the performance metric $\overline{GM}$. We fix $U=12$, $N_u=16$ and $N_b=128$. Fig.~\ref{fig:impact-beam-1} shows $\overline{GM}$ as a function of $K_b$ for different values of the number of beams at the UE and BS BF codebooks. The first observation is that increasing the number of beams at the BS or UEs improves the performance since the achieved BF gain is proportional to the number of beams. Furthermore, we can see that doubling the number of beams in the BS BF codebook is slightly more beneficial than doubling the number of beams belonging to the UE BF codebook in terms of $\overline{GM}$. The reason is that the higher number of antennas at the BS allows for realizing closer to the ideal beams and hence higher BF gains on average. Fig.~\ref{fig:impact-beam-2} depicts the percentage improvement in $\overline{GM}$ with respect to the baseline case with $2^{B_u}=4$ and $2^{B_b}=16$ as a function of $K_b$. The baseline case corresponds to the green curve with diamond markers in Fig.~\ref{fig:impact-beam-1}. An important observation from this figure is that the gain obtained by increasing the size of the codebook substantially increases with the number of RF chains $K_b$. The reason is that the BF gain is proportional to the BF codebook size\footnote{The beamwidth of each beam is inversely proportional to the number of BF codewords. Therefore, increasing the BF codebook size reduces the beamwidth for all the beams and subsequently increases the BF gain.}. Therefore, when $K_b$ is larger, a larger BF codebook can help the scheduler to exploit the increased BF gain and activate more UEs at the same PRB. We note that the increased BF gain can help the scheduler to compensate the reduced transmit power to each UE when a larger number of UEs are scheduled.  


\section{Conclusion}
We investigated multi-user scheduling in OFDM mmWave systems with per-stream hybrid BF. We formulated the scheduling as an optimization problem and provided a convex relaxed version which can be solved efficiently. Using the solution of the relaxed problem, we proposed and algorithm to build a feasible solution shown to perform very close to an upper-bound. We used the proposed solution to investigate the impact of various system parameters on the system performance and obtain various engineering insights. A natural extension to this work is to extend the system model to allow for multiple data streams per UE. Another future avenue is to study power distribution optimization among the streams.

\bibliographystyle{IEEEtran}
\bibliography{bibliography.bib}

\end{document}